\documentclass[twocolumn,english,aps,prl,showpacs]{revtex4}
\usepackage[T1]{fontenc}
\usepackage{amsmath,graphicx,amssymb,epsfig,babel,dsfont}

\renewcommand{\Im}{{\rm Im}}
\newcommand{\Tr}{{\rm Tr}}
\newcommand{\rd}{{\rm d}}
\newcommand{\kb}{k_{\rm B}}

\begin{document}

\title{Photon thermal Hall effect}

\author{P. Ben-Abdallah}
\email{pba@institutoptique.fr} 
\affiliation{Laboratoire Charles Fabry,UMR 8501, Institut d'Optique, CNRS, Universit\'{e} Paris-Sud 11,
2, Avenue Augustin Fresnel, 91127 Palaiseau Cedex, France.}


\date{\today}

\pacs{44.40.+a, 78.20.N-, 03.50.De, 66.70.-f}
\begin{abstract}
A near-field thermal Hall  effect (i.e.Righi-Leduc effect) in networks of magneto-optical particles placed in a constant magnetic field is predicted. This many-body effect is related to a symmetry breaking in the system induced by the magnetic field which gives rise to preferential channels for the heat-transport by near-field interaction thanks to the particles anisotropy tuning. 
\end{abstract}

\maketitle

The Righi-Leduc effect~\cite{Leduc} is the thermal analog of classical Hall effect~\cite{Hall}. It consists in the appearance of a heat flux transversally to a heat current induced by a temperature gradient inside a solid under the presence of a magnetic field. Like the Hall effect, it is due to the curvature of carriers trajectories through the magnetic field. At macroscopic scale this effect is related to a symmetry breaking in the transport equations due to  the presence of an external magnetic field. At microscale numerous mechanisms can be responsible for this effect.  In semiconductors, metals or high-Tc superconductors it is the Lorentz force acting on the free electrons which is responsible for a transversal heat current. In ferromagnetic materials, magnons (spin waves)~\cite{Fujimoto,Katsura,Onose} currents have been shown to be the source of thermal Hall effect. Recently, a phonon mediated themal Hall effect~\cite{Strohm,Inyushkin}  has been highlighted in neutral objects of zero electrical charge. But, so far, no magnetotransverse effect have been predicted for the photon contribution of the thermal conductivity. In this Letter, we investigate  the  near-field heat exchanges in a four-terminal  system composed by magneto-optical particles under the action of a constant magnetic field and we demonstrate the existence of a Hall flux in the direction perpendicular to the primary temperature gradient which is due to a breakdown of the symmetry  in the near-field interactions.

\begin{figure}[Hhbt]
\centering
\includegraphics[angle=0,scale=0.3]{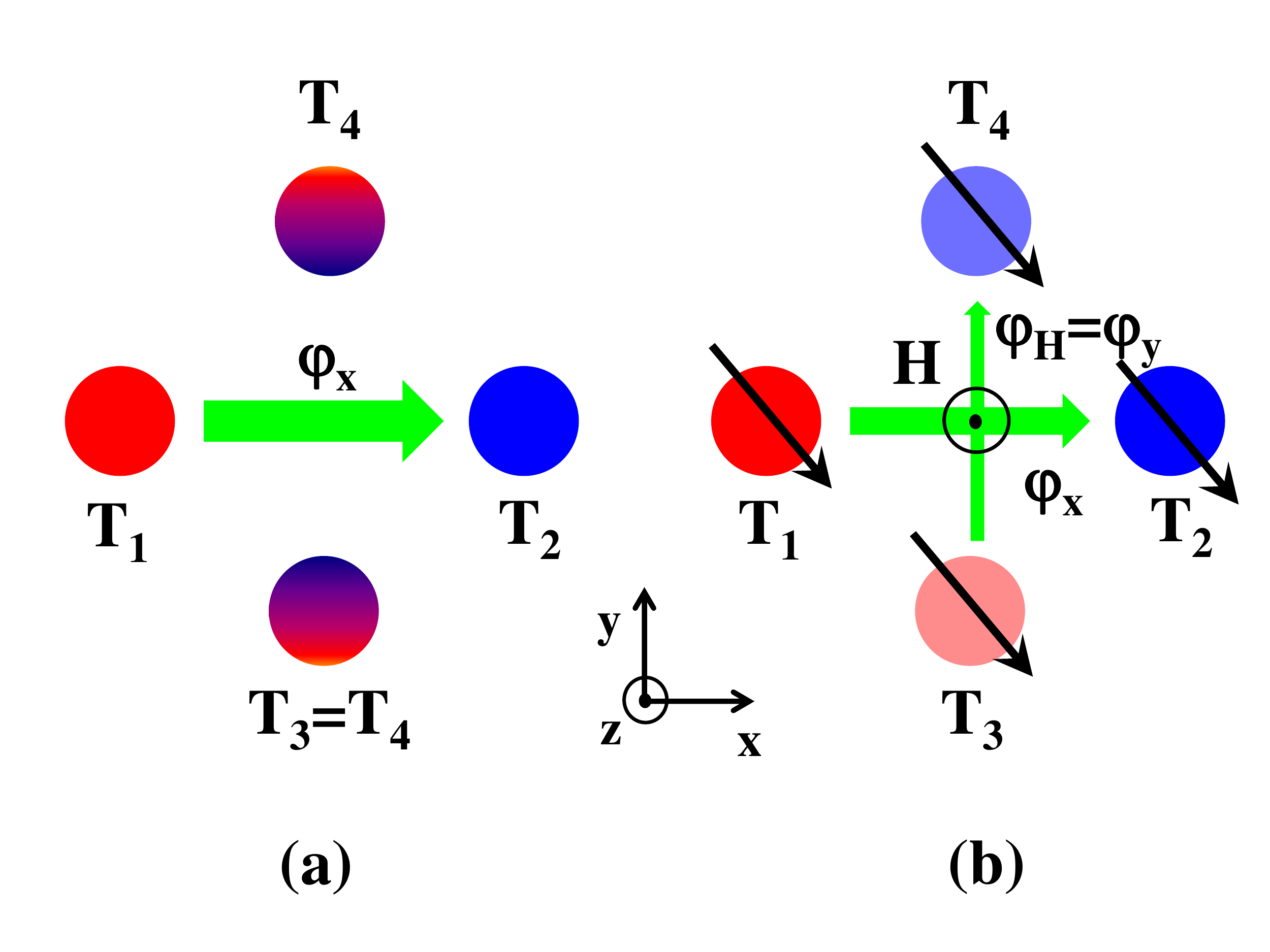}
\caption{Sketch of the four terminal junction made with magneto-optical nanoparticles used to demonstrate the existence of a  photon Hall effect under the action of an external magnetic field $\mathbf H$ when the left and right particles are held at two different temperatures. (a) If $H=0$ the particles are optically isotropes so that the system is thermally symmetric (i.e. $T_3=T_4$) and the Hall flux $\varphi_H\equiv \varphi_y$ is null. (b) If a magnetic field is applied in the $z$ direction, the particles become biaxial breaking the system symmetry (the optical axis are two complex valued vectors  $V_1=(i,1)$ and $V_2=(-i,1)$, the eigenvectors of permittivity tensor) a temperature gradient is generated in the $y$ direction giving rise to a non null Hall flux. The black arrows illustrate the particles anisotropy.} 
\end{figure}

To start, we consider the system sketched in Fig.~1. It consists in four identical spherical particles made with a magneto-optical material which are arranged in a four-terminal junction. Those particles can exchange electromagnetic energy between them and with the surrounding medium which can be assimilated to a bosonic field at ambiant temperature $T_a$. By connecting the two particles along the $\mathbf{x}$-axis to two heat baths at two different temperatures, a heat flux flows through the system between these two particles. Without external magnetic field all particles are isotropic, so that the two others unthermostated particles have, for symmetry reasons, the same equilibrium temperatures and therefore they do not exchange heat flux through the network. On the contrary, when a magnetic field is applied orthogonally to the particles network, the particles become anisotropic so that the symmetry of system is broken (Fig.~1). As we will see hereafter, when the steady state regime is reached, the two unthermostated particles display two different temperatures. Therefore a heat flux propagates transversally to the primary applied temperature gradient.

Using the Landauer formalism for N-body systems~\cite{PBAEtAl2011,Riccardo,PRL_superdiff,Nikbakht,Incardone,splitter} the heat flux exchanged between the $i^{th}$ and the $j^{th}$ particle in the network reads 
\begin{equation}
  \varphi_{ij}=\int_{0}^{\infty}\frac{\rd\omega}{2\pi}\,[\Theta(\omega,T_{i})-\Theta(\omega,T_{j})]\mathcal{T}_{i,j}(\omega)\label{Eq:InterpartHeatFlux},
\end{equation}
where $\Theta(\omega,T)={\hbar\omega}/[{e^{\frac{\hbar\omega}{k_B T}}-1}]$ is the mean energy of a harmonic oscillator in
thermal equilibrium at temperature $T$ and $\mathcal{T}_{i,j}(\omega)$ denotes the transmission coefficient, at the frequency $\omega$, between the two particles. When the particles are small enough 
compared with their thermal wavelength $\lambda_{T_{i}} = c\hbar/(\kb T_{i})$ ($c$ is the vacuum light
velocity, $2 \pi \hbar$ is Planck's constant, and $\kb$ is Boltzmann's constant) they can be modeled by simple radiating electrical dipoles.
In this case the transmission coefficient is defined as~\cite{Nikbakht}
\begin{equation}
  \mathcal{T}_{i,j}(\omega)=2\Im\Tr\bigl[\mathds{A}_{ij}\Im\bar{\bar{\boldsymbol{\chi}}}_j\mathds{C}_{ij}^{\dagger}\bigr],
\end{equation}
where $\bar{\bar{\boldsymbol{\chi}}}_j$, $\mathds{A}_{ij}$ and $\mathds{C}_{ij}$ are the susceptibility tensor plus two 
matrices which read~\cite{Nikbakht} in terms of free space Green tensor $\bar{\bar{\boldsymbol{G}}}_{ij}^{0}=\frac{\exp({\rm i}kr_{ij})}{4\pi r_{ij}}\left[\left(1+\frac{{\rm i}kr_{ij}-1}{k^{2}r_{ij}^{2}}\right)\mathds{1}+\frac{3-3{\rm i}kr_{ij}-k^{2}r_{ij}^{2}}{k^{2}r_{ij}^{2}}\widehat{\mathbf{r}}_{ij}\otimes\widehat{\mathbf{r}}_{ij}\right]$
($\widehat{\mathbf{r}}_{ij}\equiv\mathbf{r_{\mathit{ij}}}/r_{ij}$, $\mathbf{r_{\mathit{ij}}}$ is the vector linking the center of
dipoles i and j, while $r_{ij}=\mid\mathbf{r}_{ij}\mid$ and $\mathds{1}$ stands for the unit dyadic tensor) and of polarizabilities matrix $\hat{\mathds{\alpha}}=diag(\bar{\bar{\boldsymbol{\alpha}}}_{1},...,\bar{\bar{\boldsymbol{\alpha}}}_{N})$ ($\bar{\bar{\boldsymbol{\alpha}}}_{i}$ being the polarizability tensor associated to the $i^{th}$ object)
\begin{equation}
\bar{\bar{\boldsymbol{\chi}}}_j=\bar{\bar{\boldsymbol{\alpha}}}_{j}-i\frac{k^3}{6\pi} \bar{\bar{\boldsymbol{\alpha}}}_{j}\bar{\bar{\boldsymbol{\alpha}}}_{j}^{\dagger},
\end{equation}
\begin{equation}
\mathds{A}_{ij}=\left[\mathds{1}-k^2\hat{\mathds{\alpha}}\mathds{B}\right]_{ij}^{-1},
\end{equation}
with $\mathds{B}_{ij}=(1-\delta_{ij})\bar{\bar{\boldsymbol{G}}}_{ij}^{0}$ and
\begin{equation}
\mathds{C}_{ij}=k^2\bar{\bar{\boldsymbol{G}}}_{ik}^{0}\mathds{A}_{kj}.
\end{equation}
If the temperature difference between the two thermostated particles is small (i.e. $T_i=T_{eq}+\Delta T_i$, $T_{eq}$ being the temperature of cold reservoir) then we can treat the system in linear regime. In this case, the heat flux received by each particle can be written as
\begin{equation}
  \phi_{i}=\underset{j\neq i}{\sum}\varphi_{ij}=\underset{j\neq i}{\sum}G_{ij}(T_j-T_i)\label{Eq:HeatFlux},
\end{equation}
where 
\begin{equation}
 G_{ij}=\frac{\partial\varphi_{ij}}{\partial T}\mid_{T=T_{eq}}=\int_{0}^{\infty}\frac{\rd\omega}{2\pi}\frac{\partial\Theta}{\partial T}\mathcal{T}_{i,j}(\omega)\label{Eq:Conductance}
\end{equation}
is the thermal exchange conductance between the  $i^{th}$ and the $j^{th}$ particle at temperature $T_{eq}$. In steady state, the net power received by each particle vanishes. By neglecting the far field interactions with the surrounding field (the power $\varphi_{i\leftrightarrow a}=\overline{C}_{\text{abs};i}\sigma_{B}(T_\text{a}^{4}-T_i^{4})$ exchanged in far field with the environment where $\overline{C}_{\text{abs};i}$ is the thermally averaged dressed absorption cross-section of the $i$-th particle and $\sigma_B$ is the Stefan-Boltzmann constant is negligeable in front of near-field interactions~\cite{Yannopapas}) the energy balance equation reads
\begin{align}
\mathcal{P}_{i,b_i}+\underset{j\neq i}{\sum}G_{ij}(T_j-T_i)=0, \quad i=1,2 \\
\underset{j\neq i}{\sum}G_{ij}(T_j-T_i)=0, \qquad i=3,4,
\end{align}
where $\mathcal{P}_{i,b_i}$ is the power exchanged between the thermostated particle $i$ and the $i^{th}$ heat bath. By solving the two last equations with respect to unknown temperatures $T_3$ and $T_4$ we get
\begin{equation}
\begin{split}
T_3=\frac{1}{\Upsilon}[(G_{31}\underset{j\neq 4}{\sum}G_{4j}+G_{34}G_{41})T_1)
\\+(G_{32}\underset{j\neq 4}{\sum}G_{4j}+G_{34}G_{42})T_2)]\label{Eq:T3}, 
\end{split}
\end{equation}
\begin{equation}
\begin{split}
T_4=\frac{1}{\Upsilon}[(G_{41}\underset{j\neq 3}{\sum}G_{3j}+G_{43}G_{31})T_1)
\\+(G_{42}\underset{j\neq 3}{\sum}G_{3j}+G_{43}G_{32})T_2)]\label{Eq:T4}
\end{split}
\end{equation}
where $\Upsilon=\underset{j\neq 3}{\sum}G_{3j}\underset{j\neq 4}{\sum}G_{4j}-G_{34}G_{43}$. From expressions (\ref{Eq:T3}) and (\ref{Eq:T4}) and using the reciprocity of heat exchanges (i.e. $G_{ij}=G_{ji}$) we find the condition to fulfilled in order to get a null Hall flux (i.e. $T_3=T_4$)
\begin{align}
G_{31}G_{42}-G_{41}G_{32}=0\label{Eq:Cond_PHE}. 
\end{align}
Thus, by breaking this symmetry condition into the system applying, for instance, an external magnetic field, a temperature difference must appear between the upper and lower particles giving rise to a thermal Hall flux.

The magnitude of this Hall effect can be evaluated using the relative Hall temperature difference
\begin{equation}
 R=\frac{T_3-T_4}{T_1-T_2} \label{Eq:relative1}.
\end{equation}
In linear response regime, this expression reads from relations (\ref{Eq:T3}) and (\ref{Eq:T4})
\begin{equation}
 R=\frac{G_{31}G_{42}-G_{41}G_{32}}{\Upsilon} \label{Eq:relative2}.
\end{equation}

\begin{figure}[Hhbt]
\centering
\includegraphics[angle=0,scale=0.26,angle=0]{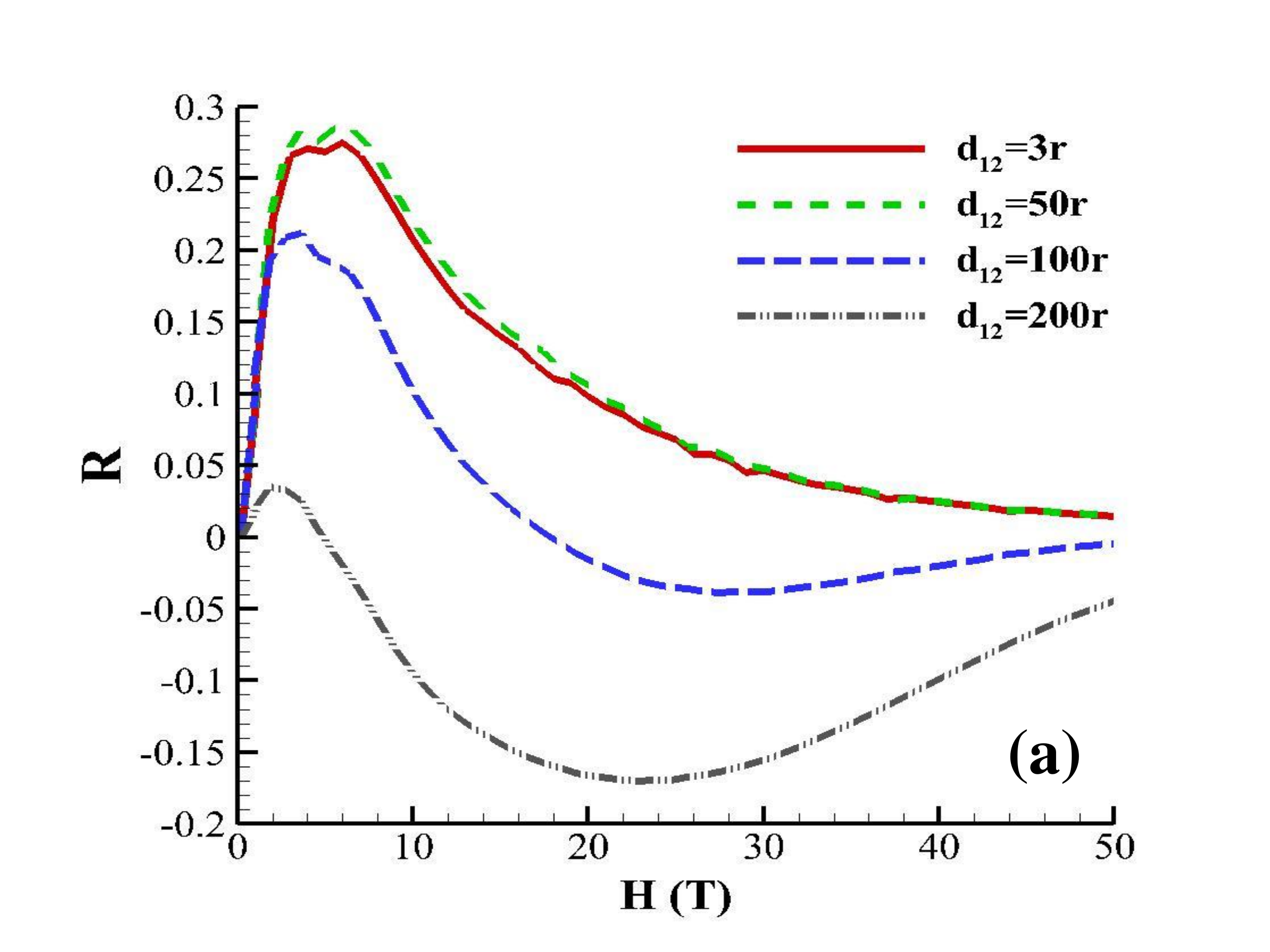}
\includegraphics[angle=0,scale=0.26,angle=0]{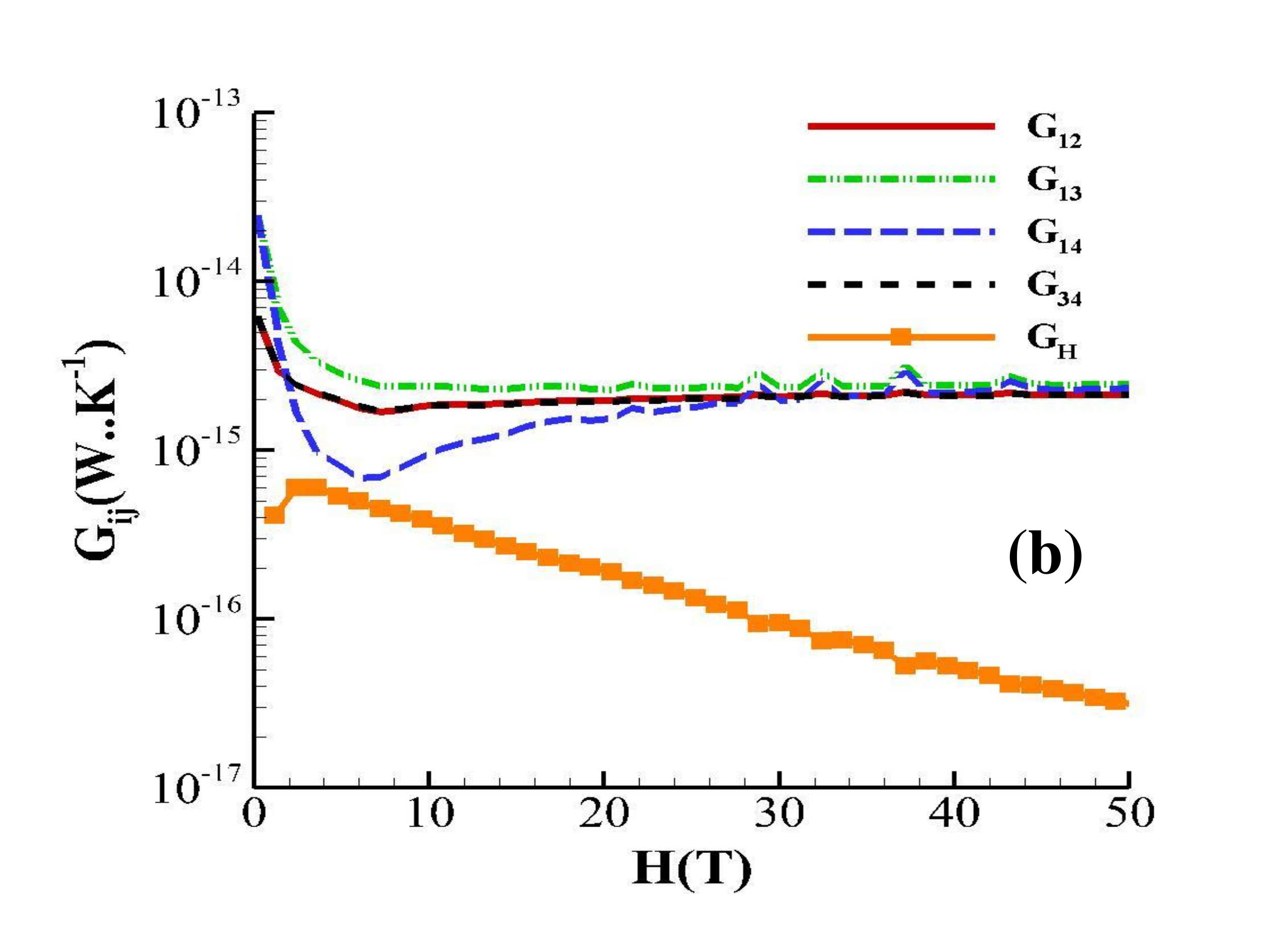}
\includegraphics[angle=0,scale=0.25,angle=0]{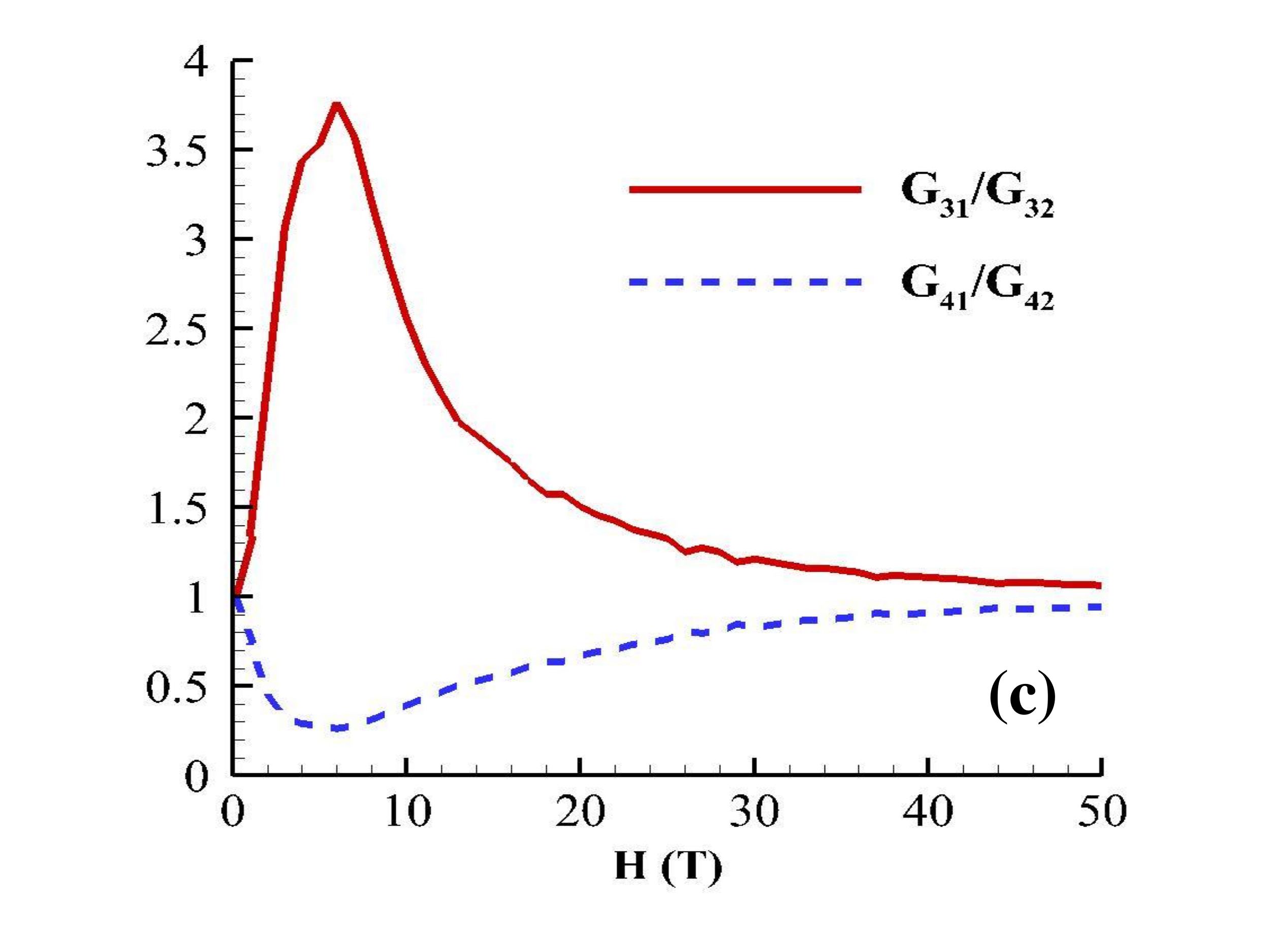}
\caption{(a) Relative Hall temperature difference $R$ versus the magnetic field $H$ in a four-terminal square junction of InSb particles $r=100 nm$ radius at $T_{eq}=300 K$. The separation distance $d_{12}$ (from edge to edge) is $3 r, 50r$ (near-field), $200r$ (far-field) and $100r$ (intermediate regime). (b) Thermal exchange conductances  $G_{12}$, $G_{13}$, $G_{14}$, $G_{34}$ and Hall conductance $G_H$ at $T_{eq}= 300 K$ with respect to the magnetic field intensity when $d_{12}=d_{34}=3r$. (a) Asymmetry factors of conductances at $T_{eq}= 300 K$ when $d_{12}=3r$.}
\end{figure}
As for the Hall conductance, it is defined from the ratio of Hall flux 
\begin{equation}
\varphi_H=\underset{i\neq4}{\sum}\varphi_{4i}-\underset{i\neq3}{\sum}\varphi_{3i}\label{Eq:Hall flux}
\end{equation}
over the primary temperature gradient $T_1-T_2$ when $T_{1}\rightarrow T_{2}$. After a straighforward calculation~\cite{SupplMat} this leads to
\begin{equation}
\begin{split}
G_H=(G_{42}-G_{41})_{T_{eq}}\quad\quad\quad\quad\quad\quad
\\-R(G_{14}+G_{24}+2G_{34})_{T_{eq}}\label{Eq:Hall conductance},
\end{split}
\end{equation}
where the symbol $()_{T}$ means that the conductances are calculated at temperature $T$. According to conditions (\ref{Eq:Cond_PHE}), when $R=0$ we verify that the Hall conductance vanishes.

Let us now consider a concrete situation by studying a four terminal  junction made with four identical InSb spherical particles of radius $r$ placed at the vertices of a square as skteched in Fig. 1. When a magnetic field is applied in the direction parallel to the $\mathbf{z}$-axis, the permittivity tensor of InSb particles takes the following form~\cite{Palik,Moncada}
\begin{equation}
\bar{\bar{\varepsilon}}=\left(\begin{array}{ccc}
\varepsilon_{1} & -i\varepsilon_{2} & 0\\
i\varepsilon_{2} & \varepsilon_{1} & 0\\
0 & 0 & \varepsilon_{3}
\end{array}\right)\label{Eq:permittivity}
\end{equation}
with
\begin{equation}
\varepsilon_{1}(H)=\varepsilon_\infty(1+\frac{\omega_L^2-\omega_T^2}{\omega_T^2-\omega^2-i\Gamma\omega}+\frac{\omega_p^2(\omega+i\gamma)}{\omega[\omega_c^2-(\omega+i\gamma)^2]}) \label{Eq:permittivity1},
\end{equation}
\begin{equation}
\varepsilon_{2}(H)=\frac{\varepsilon_\infty\omega_p^2\omega_c}{\omega[(\omega+i\gamma)^2-\omega_c^2]}\label{Eq:permittivity2},
\end{equation}
\begin{equation}
\varepsilon_{3}=\varepsilon_\infty(1+\frac{\omega_L^2-\omega_T^2}{\omega_T^2-\omega^2-i\Gamma\omega}-\frac{\omega_p^2}{\omega(\omega+i\gamma)})\label{Eq:permittivity3}.
\end{equation}
Here, $\varepsilon_\infty=15.7$ is the infinite-frequency dielectric constant, $\omega_L=3.62\times10^{13} rad.s^{-1}$ is the longitudinal opical phonon frequency, $\omega_T=3.39\times10^{13} rad.s^{-1}$ is the transverse optical phonon frequency, $\omega_p=(\frac{ne^2}{m^*\varepsilon_0\varepsilon_\infty})^{1/2}$ is the plasma frequency of free carriers of density $n=1.07\times10^{17} cm^{-3}$ and effective mass $m^*=1.99\times 10^{-32}kg$, $\Gamma=5.65\times10^{11} rad.s^{-1}$ is the phonon damping constant,$\gamma=3.39\times10^{12} rad.s^{-1}$ is the free carrier damping constant and $\omega_c=eH/m^*$ is the cyclotron frequency. Thus, the polarizability tensor for a spherical particles can be described, including the radiative corrections, by the following anisotropic polarizability~\cite{Albaladejo}
\begin{equation}
\bar{\bar{\boldsymbol{\alpha}}}_{i}(\omega)=( \bar{\bar{\boldsymbol{1}}}-i\frac{k^3}{6\pi} \bar{\bar{\boldsymbol{\alpha_0}}}_{i})^{-1} \bar{\bar{\boldsymbol{\alpha_0}}}_{i}\label{Eq:Polarizability},
\end{equation}
where $ \bar{\bar{\boldsymbol{\alpha_0}}}_{i}$ denotes the quasistatic polarizability of the $i^{th}$ particle which reads for spheres made with magneto-optical materials and which are embedded inside an isotropic host of permittivity $\varepsilon_h$
\begin{equation}
  \bar{\bar{\boldsymbol{\alpha_0}}}_{i}(\omega)=4\pi r^3(\bar{\bar{\varepsilon}}-\varepsilon_h\bar{\bar{1}})(\bar{\bar{\varepsilon}}+2\varepsilon_h\bar{\bar{1}})^{-1}\label{Eq:Polarizability2}.
\end{equation}
As shown in the supplementary material~\cite{SupplMat} the particles polarizability becomes strongly anisotropic in presence of magnetic field. It is also shown, for particles smaller than the wavelength, that the contribution of  magnetic moments can be neglected in front of electric contributions in the dissipation process.

In Fig. 2-a we show the relative Hall temperature difference $R$  with respect to the magnitude $H$ of magnetic field both in near-field and far-field regimes when the particles are embedded in vacuum (i.e. $\epsilon_h=1$). 
For any separation distance, when the magnetic field is zero, all particles are isotropic so that the system is symmetric and, as expected, $R=0$. On the contrary, for non null magnetic field the symmetry of system is broken and a Hall flux appears. The results plotted in Fig. 2-a show that, in near-field regime, $R$ keeps the same sign whatever the magnitude of magnetic field. On the contrary, in far-field regime we see that the sign of $R$ can changes with the magnitude of magnetic field. Since, the optical properties of particles are the same for both regimes of interaction, this difference of behavior comes from the spatial variation of electric field itself radiated by each particle. Under the action of a weak magnetic field, the spatial distribution of electric field radiated by the particles leads to a strongest dissipation of energy in the lower particle (particle 3) so that a Hall flux flows in the direction of positives $y$.  On the other hand, for strong magnetic fields of magnitude larger than about $H=5 T$  it is the upper particle (particle 4) which is over excited and the Hall flux goes in the opposite direction. However, as shown in Fig. 2-a,  in far-field regime, the difference of temperature between particles 3 and 4 is generally much smaller in far-field than in near-field regime for weak field. This comes from the efficiency of heat exchanges in near-field because of the presence of surface waves. Hence, as $d_{12}=3r$, the Hall temperature difference is approximatly equal to $28\%$ of the primary temperature gradient when $H=3 T$ while it is equal to about $\%3$ in far-field regime. Because of this, hereafter we focus  exclusively our attention on the near-field regime.

To go further in the thermal Hall effect analyzis, let us examinate now the variation of thermal conductances with the magnitude of magnetic field. For low field, we observe that $G_{13}$ and $G_{14}$ are notably different. This difference gives rise to a preferential channel for  heat echanges through the network. Around  $H=6T$ the asymmetry inside the system becomes maximal as shown in Fig. 2-c so that $R$ becomes maximal as well. When $G_{13}>G_{14}$, a larger amount of energy is transmitted from the first (hot) particle to the third particle than from the first particle to the fourth. Therefore, the third particle  becomes hotter than the fourth one and accordingly a Hall flux flows toward the latter (i.e. $R>0$).
To explain the near-field coupling between the particles, let us focus now our attention on the optical properties of particles.  According to the Clausius-Mosotti-like relation (\ref{Eq:Polarizability2}), the particle resonances, which correspond to localized surface polaritons, are  solutions of the following transcendental equations
\begin{equation}
 \varepsilon_{3}(\omega)+2\varepsilon_{h}=0\label{Eq:res1}
\end{equation}
and
\begin{equation}
( \varepsilon_{1}(\omega)+2\varepsilon_{h})^2-\varepsilon_{2}^2(\omega)=0\label{Eq:res2}.
\end{equation}

\begin{figure}[Hhbt]
\centering
\includegraphics[angle=0,scale=0.25]{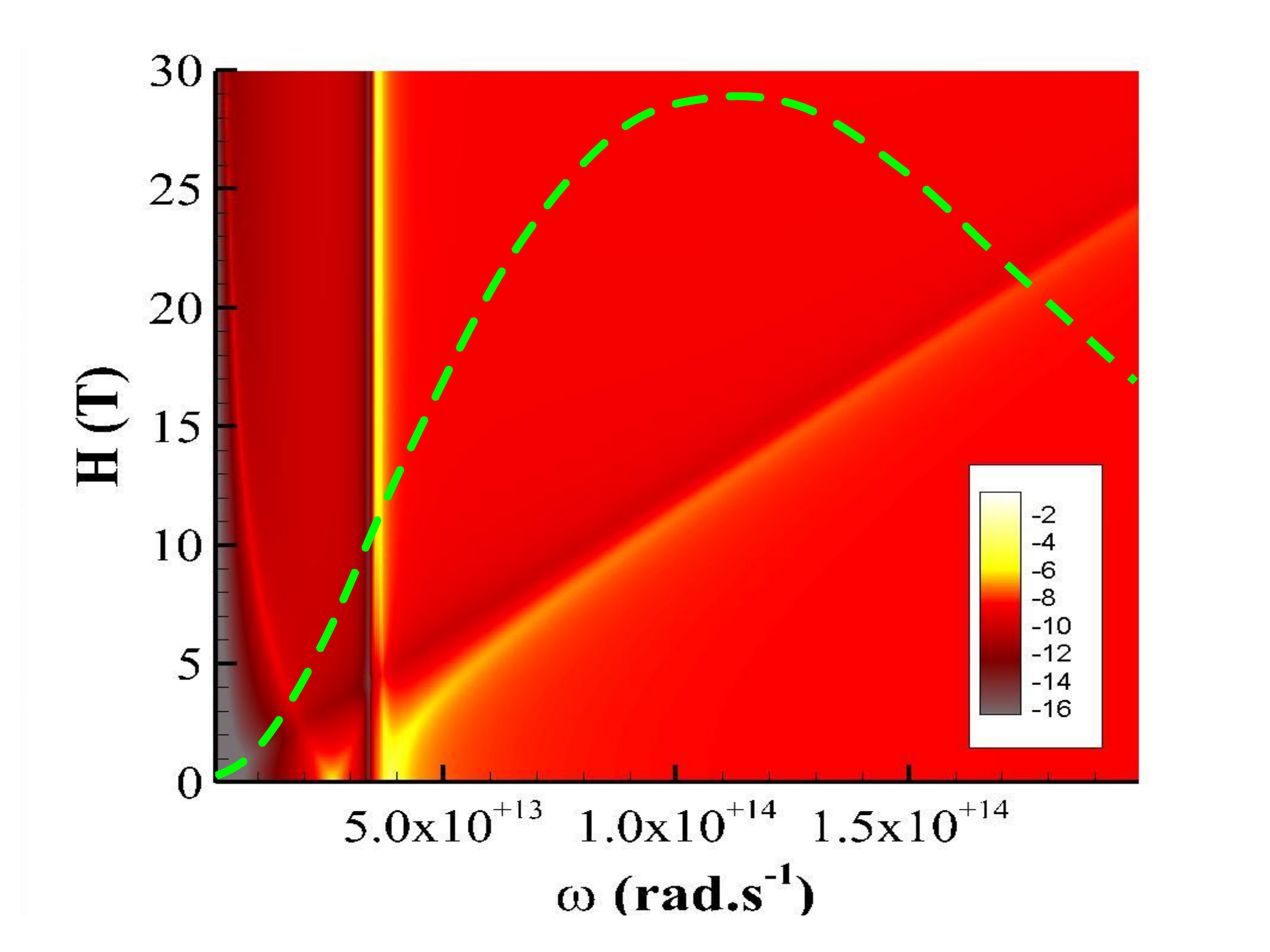}

\caption{Resonance conditions of  InSb particles in vacuum (i.e. $\epsilon_h=1$). (a) Plot of $ln(\mid(\epsilon_3+2\epsilon_h)[(\epsilon_1+2\epsilon_h)^2-\epsilon^2_2]\mid^{-1})$ in the $(\omega,H)$ plane. The dashed line corresponds to the Planck function (arbitrary unit) at $T_{eq}=300 K$.}
\end{figure}

These resonances are plotted in Fig. 3 with respect to the magnitude of magnetic field.  We clearly see the presence of three different branches (bright areas). The vertical branch  is independent on the magnetic field. This branch is related to the resonance which is solution of Eq. (\ref{Eq:res1}) and it corresponds to the presence of a surface phonon polariton (SPhP) at $\omega\sim 3.5\times10^{13} rad.s^{-1}$. The two others branches are solutions of Eq. (\ref{Eq:res2}). Contrary to the first resonance, these resonances depend on the magnetic field and are of plasmonic nature. When the magnitude of magnetic field becomes sufficiently large, these plasmonic resonances get away from the Wien's frequency so that they do not contribute anymore to  heat exchanges. On the other hand, for weak magnetic fields, these resonances give rise to supplementary channels for heat exchanges which superimpose to the channel associated with SPhP.  Moreover, the contribution of the high frequency plasmonic channel becomes more and more important as its frequency brings closer from the Wien's frequency. The optimal transfer occurs for a magnetic field of magnitude $H=6 T$.  This situation corresponds precisely to the condition where the Hall effect is maximal.

If an experimental observation of photon thermal Hall effect seems to be out of reach with nanoparticle networks, a direct measurement of the Hall temperature difference with measurements of electrical resistance variations in magneto-optical nanowires networks should be feasible. A similar experiment has been reported recently~\cite{Lipson} to measure, with a very high occuracy, the near-field heat exchanges between two nanobeams. 
Beside the experimental observation of this effect, some potential applications of photon thermal Hall effect may be considered. As for the classical Hall sensor, a many body junction made with magneto-optical elements is a natural  building block to make a purely thermal magnetic field detection. Indeed, in linear regime, the Hall flux $\varphi_H=G_H R \Delta T$ ($\Delta T$ being the primary temperature gradient) is directly proportional through the Hall conductance to the magnetic field. Another application is the use in nanoscale heat engines of thermal Hall effect in presence of an AC magnetic field to modulate the heat flow in multiple directions. Of course, the upper frequency for such a modulation is limited by the thermal relaxation of the Hall cell. But, with nanocomponents, this frequency is of the order of the inverse of phonon relaxation time $\tau_{ph}\sim 10^{-12}s$.

\begin{acknowledgements}
P.B.-A. acknowledges discussions with S.-A. Biehs, R. Esquivel and D. Dalvit.
\end{acknowledgements}

\end{document}